\documentstyle[twocolumn,aps,epsfig,prb]{revtex}
\begin{document}
\draft
\twocolumn[\hsize\textwidth\columnwidth\hsize\csname@twocolumnfalse\endcsname

\title{Carbon Nanotube - Substrate Interface Interactions }

\author{ R. Czerw$^1$, B. Foley$^{1,2}$, D. Tekleab$^1$, A. Rubio$^3$, P. M. Ajayan$^4$, and D. L. Carroll$^1$}

\address{
             $^1$Dept. of Physics and Astronomy, Clemson University, Clemson SC 29634-0978, USA\\
             $^2$Dept. of Physics, Trinity College Dublin, Dublin 2, Ireland\\
             $^3$Departamento de Fisica de Materiales, UPV/EHU, Donostia International Physics Center (DIPC),
             Centro Mixto CSIC-UPV/EHU, 20018 San Sebastian, Spain\\
             $^4$Dept. of Materials Science and Engineering, Rensselaer Polytechnic Institute, Troy NY 12180-3590,
             USA}

\date{\today}
\maketitle

\begin{abstract}
By utilizing the current transients in scanning tunneling
spectroscopy, the local interfacial electronics between
multiwalled carbon nanotubes and several supporting substrates has
been investigated.  Voltage offsets in the tunneling spectra are
directly correlated with the formation of a dipole layer at the
nanotube-substrate interface strongly suggesting the formation of
interface states.  Further, a systematic variation in this local
potential, as a function of tube diameter, is observed for both
metallic substrates (Au) and semi-metallic substrates (graphite).
In both cases, tubes with diameters between $\sim$ 5 nm and 30 nm,
the interfacial potential is nearly constant as a function of tube
diameter.  However, for tube diameters $<$ 5 nm, a dramatic change
in the local potential is observed.  Using {\it ab initio}
techniques, this diameter dependent electronic interaction is
shown to derive from changes in the tube-substrate hybridization
that results from the curvature of the nanotubes.
\end{abstract}

\vskip1pc]

Over the last several years, an astonishing number of experimental
determinations of the electronic transport properties of
single-walled carbon nanotubes (SWNTs) and multiwalled carbon
nanotubes (MWNTs) have been made \cite{dek99,chico96,liu01}  and
correlated with a number of theoretical predictions.\cite{rubio96}
 With only a few exceptions, the experimental focus has been on
nanotubes supported on a substrate of some type, and in all cases
a system of contacts has been employed using traditional
interconnect materials such as Cu, Au, Pt, and Ag. Thus, all the
experimental information we currently have on the electronics of
carbon nanotubes involves experimental designs that use
metal/semiconductor-nanotube interfaces and in most cases these
interfaces include the length of the tube through the support.
Clearly an important part of any interpretation of transport
results must include interactions that may exist between the
nanotube and the support substrate / contacts.\cite{tans98} Recent
theoretical studies have addressed Fermi-level alignment in
Au-SWNT systems and suggested that a charge transfer should exist
at the interface. \cite{xue99}  Similarly, some tunneling
spectroscopy experiments have hinted at the existence of charge
transfer between the gold and nanotube systems. \cite{ven99}
However, despite the importance of local interface interactions in
transport measurements, no direct determinations of its variation
with tube diameter, tube chirality, etc. is yet available.  In
this letter we present an investigation of the interfacial
electronic structure between MWNTs and a number of support
substrates using current transients in scanning tunneling
spectroscopy (STS). These studies strongly suggest that charge
transfer at the metal/MWNT interface results from the formation of
interface states in analogy to bulk Schottky barriers.  Further,
the variation of local interface potentials with tube diameter is
nearly constant as expected for large diameter tubes.  However, a
surprisingly large variation in interface potentials occurs at
tube diameters smaller than approximately 5 nm.  These variations
results from differences in local hybridization at the
metal-nanotube interface caused by the different curvatures of the
nanotube walls.

Measurements of the contact potential of three support substrates
will be described and compared here; NiO, Au, and HOPG (highly
oriented pyrolytic graphite). In the first case, the NiO
substrates are thin oxide films (5.0 nm) thermally grown on single
crystal (111) Ni.  In the second case, Au was sputter deposited on
annealed Mica and then ``flamed'' leaving a terraced (111) surface
for contact with the tubes. STM imaging of these substrates
yielded atomic resolution.  After substrate preparation, arc grown
MWNTs,\cite{ebb92} ultrasonically dispersed in ethanol for 5
minutes, were deposited. The sample was transferred into UHV
(ultra high vacuum, $<$ 10 $^{-10}$ torr) and out-gassed at 300
$^\circ$C while adsorbate desorption was monitored to insure that
the substrate was clean. Tunneling conditions used for STM imaging
were 20 pA and between 200 to 500 mV.  In the case of HOPG
substrates, HOPG was cleaved in air and an ultrasonicated solution
of nanotubes in ethanol was deposited as above. STM imaging at 20
pA and 300 mV allowed atomic resolution of the substrate and of
the supported tubes. Transmission electron microscopy images were
correlated to STM micrographs to insure the same distribution of
diameters of the MWNTs was observed (approximately 2 nm to 30 nm).
Z-scale calibration of the STM was carried out using known step
heights of the Au (111) surface. Finally, work function
determinations were compared to the reported values over the clean
substrates as a check of Z calibration.

\begin{figure}[!t]
\centerline{\epsfig{figure=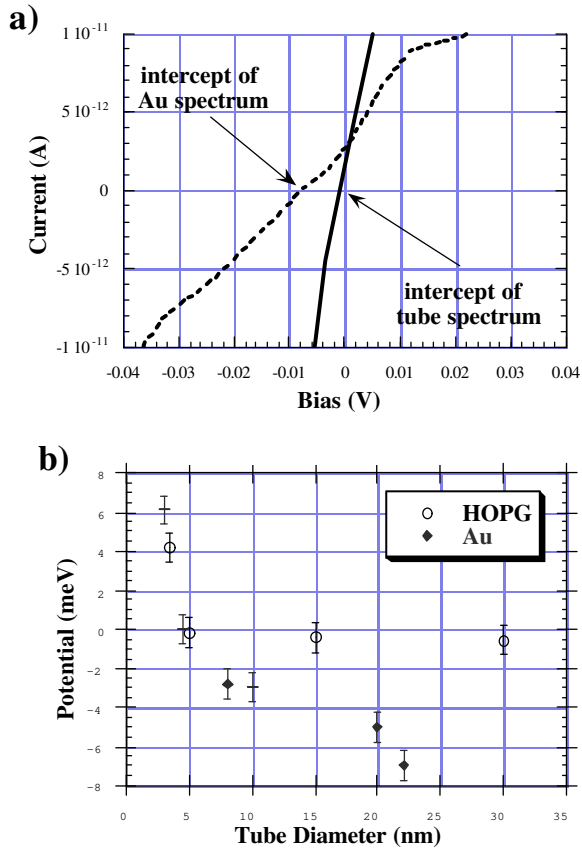,width=1.0\linewidth,clip}}\vspace{1
pc} \caption{(a) Tunneling spectra were collected over the
substrate and over the tube for each measurement made. The offsets
in tunneling spectra are due to capacitance of the junction plus
local potentials within the junction.  Subtracting the shift
observed for the IV curve on the tube from that observed over from
the substrate should give zero unless there are extra potentials
within the system. (b) The contact potential as determined from
the offset in the ``high ramp rate'' tunneling spectra is strongly
dependent on the diameter of the nanotube as shown here.  The
Au-nanotube samples and the HOPG-nanotube samples are shown.}
\label{fig1}
\end{figure}

To investigate the nanotube-substrate interface, tunneling spectra
(IV) were acquired both on the clean support substrate and at
points on the tube, simultaneously with imaging.  In our case, the
spectra are acquired by turning off the feed back over the point
of interest, ramping the voltage very rapidly, and collecting the
current.  The rapid ramp rate of the voltage results in current
transients that offset the tunneling spectra as shown in Fig.\
\ref{fig1}a. The RC time constants associated with capacitances,
contact potentials, filter, etc. in the tunneling microscope are
very short compared to those of the tunneling
junction.\cite{short} Therefore, the transient-induced shift in
the IV spectra can be associated only with capacitances and
potentials within the tunneling junction. The current-axis
intercept point (V=0) is well known to be related to the
capacitance of the junction from the equation q= VC, giving dq/dt
= CdV/dt + VdC/dt = CdV/dt.  The point at which the spectrum
crosses the V axis (I=0) is the related to the local potentials
across the junction and corresponds to the bias offset for which
the microscope must compensate to prevent the ``discharge'' of the
junction.  This includes the voltage drop across the tip-sample
capacitor, as well as any local potentials such as the contact
potentials.\cite{ind} The spectra and intercepts for a 22.0 nm
diameter tube supported on Au are shown in Fig.\ \ref{fig1}a.

\begin{figure}[!t]
\centerline{\epsfig{figure=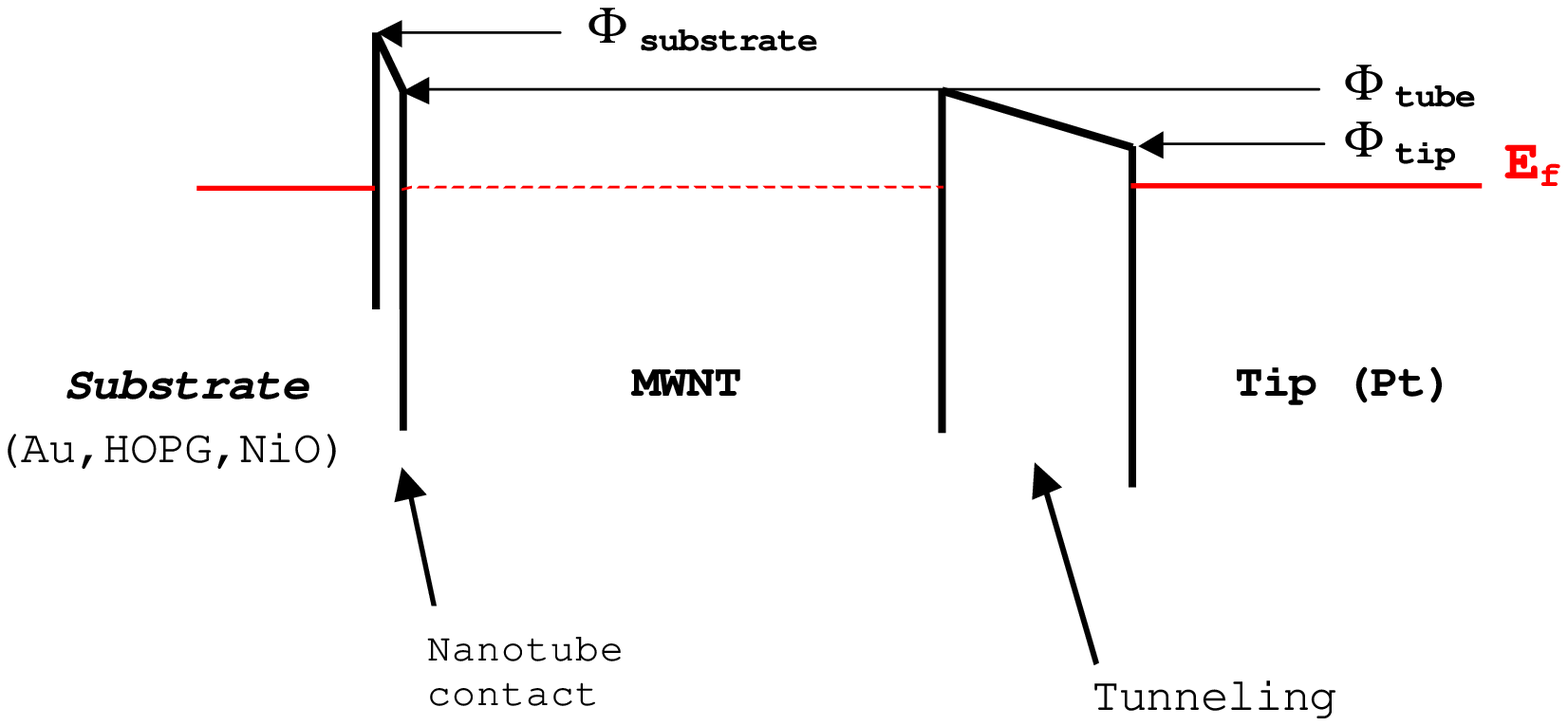,width=1.0\linewidth,clip}}\vspace{1
pc} \caption{An energy diagram of the substrate-nanotube-tip
system.  In the ideal case where no interface state exists, there
should be a simple alignment of Fermi levels and the tube
contributions to the potentials within the junction should
cancel.} \label{fig2}
\end{figure}

Naturally, we would expect that the comparatively large potentials
associated with the capacitive junction and the work function
mismatch of the substrate and the Pt tip would dominate the IV
shift.  That is, when a nanotube is placed in the tunneling
junction it would have little effect outside of changing the
junction capacitance through geometry (which would be negligible).
This is illustrated in the energy diagram of Fig.\ \ref{fig2}.
Clearly, the contact potentials within the junction-nanotube
system will cancel the nanotube component as: ($\phi_{substrate}$
- $\phi_{tube}$) + ($\phi_{tube}$ - $\phi_{tip}$) =
$\phi_{substrate}$ - $\phi_{tip}$. However, notice that the
intercepts in Fig.\ \ref{fig1}a for the clean Au surface and the
nanotube-Au system are significantly different.\cite{we} In fact,
the same is true for the NiO-MWNT system using 20 nm diameter
nanotubes.  Specifically, in the Au-nanotube system, with a 10 ms
voltage step time (the step width during the voltage ramp) and a 3
ms sample delay time (see reference 11), a 20 nm nanotube will
exhibit a difference in V-intercept shift of 0.075 from that
measured on the clean substrate (just subtracting the two
intercepts).  In the NiO system, using the same ramping
conditions, the shift difference is 0.170 V!  For HOPG substrates,
large diameter nanotubes give the same value of the voltage offset
as the clean substrate.  This suggests that there exists local
trapped charge at the Au and NiO - nanotube interfaces while at
the HOPG - nanotube interface there are no extra potentials. Using
an analogy to bulk contacts, these interfaces then behave as
though there are interface states formed between the nanotubes and
the substrates (the rapidly ramped IV is somewhat analogous to
capacitance voltage  (CV) curves).\cite{rho87} No such interface
states appear in the HOPG-nanotube system for large diameter tubes
as would be expected since this looks very much like graphite on
graphite.

Twenty tubes supported by Au were studied in this way with
diameters that ranged from approximately 3 nm to 30 nm.  The step
width used in the voltage ramp was 100 ms and the delay time was 5
ms.  The difference in voltage shifts from the support substrate
was determined for these tubes by the method above (subtracting
the voltage intercept with the tip over the clean substrate from
the intercept with the tip over the tube) and is shown in Fig.\
\ref{fig1}b as a function of tube diameter.  Notice that from
approximately 10 nm to 30 nm diameter, this difference is
practically constant. Between 5 and 10 nm diameters, there is a
strong variation in the difference in voltage intercepts.  Below 5
nm the differences in the intercepts actually reverses sign!
Again, we note that this measurement gives the relative variation
in the local potential at the nanotube - substrate interface.
Clearly, as the nanotube diameter changes, the local potential
varies sharply and actually reverses sign at ~ 5 nm.  Fig.\
\ref{fig1}b also shows similar measurements performed on HOPG
substrates. While the overall difference in the voltage intercepts
is 0 V for large nanotubes as expected, a sharp variation in the
local potential is observed at around 5 nm also. On HOPG, fewer
tubes were studied (six), and we note that the values for the
larger tubes are not exactly 0 eV as would be expected (graphite
on graphite) due to the slight difference in the capacitance
between the tip-substrate and the tip-nanotube-substrate
junctions.

To understand this behavior, there are two geometrical factors
that must be estimated.  The first is the variation in the contact
area of the tube to the substrate.  This changes with tube
diameter and will alter the total charge at the interface in
direct proportion to the diameter.   A simple estimate of the
change in contact area requires a guess as to how close the tube
wall must be to be considered in contact.  We label this parameter
``h''.  From simple geometrical arguments we get: area of
contact/(unit length) = (2R){cos-1 ((R-h)/R)}.  ``h'' must be of
the order of the pz orbitals (0.1 nm).  The second geometrical
factor to consider is the effect of the tube diameter on junction
capacitance.  The capacitance change for the tip-tube-substrate
geometry was calculated using classical electrostatics.  The total
variation in junction capacitance was smoothly varying for tubes
of the diameter used in our measurements and also quite small.
Thus, we assume in this measurement that this effect is nearly
constant. While these geometrical factors could clearly account
for the smoothly varying potential associated with large tube
diameters, they fail to explain the drastic changes observed below
5 nm. Further, these can not account for the observed sign changes
in the potential. This suggests that the local charge at the
nanotube-substrate interface varies strongly with tube diameter.
Further, it is likely that these variations result from changes in
the interface electronic structure such as hybridization shifts,
shifts in local orbital occupancy, or interlayer interactions and
not geometric factors.

\begin{figure}[!t]
\centerline{\epsfig{figure=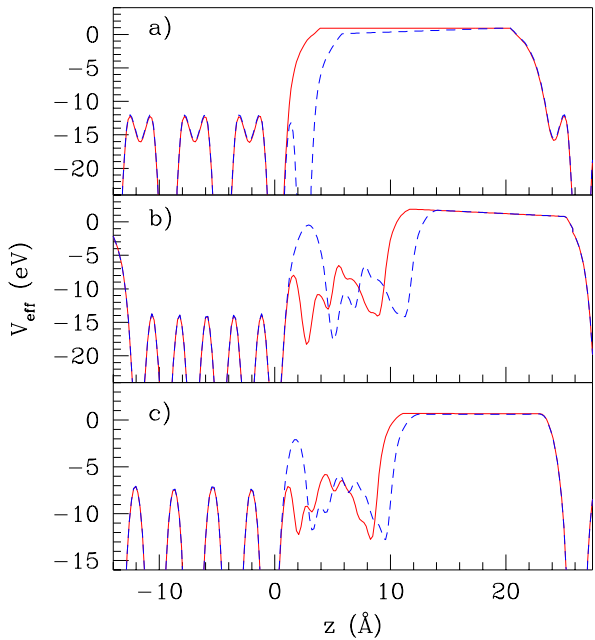,width=1\linewidth,clip}}\vspace{1
pc} \caption{{\it ab initio} calculations of the effective
electron potential in a supercell geometry for a graphene layer
and a C(5,5) tube on Au(111) a) and b), respectively, and C(5,5)
on graphite, c).  In all cases we plot the effective potential
along a line perpendicular to the gold surface.  Due to
computation limitations we have only considered six gold layers in
b). The results in a) correspond to the isolated Au(111) surface
(continuous line) and a single graphene sheet supported on it
(dashed line). The formation of a dipole barrier due to the
difference in work functions of graphite and gold is clear.  When
a small diameter single-wall nanotube is supported on Au(111) (b)
there is a clear change of sign of the  barrier. We have plotted
the effective potential for two distances, 0.22 nm (continuous
line) and 0.34 nm (dashed line), of the tube with respect to
Au(111).  This is directly connected with the observed change of
sign in the interface potential in Fig.\ \ref{fig1}b.  In c) we
plot the same as in b) but this time for a tube supported on
graphite. Notice here, there is no dipolar barrier formed.}
\label{fig3}
\end{figure}

The results in Fig.\ \ref{fig1}b point to the existence of trapped
charge at the nanotube-substrate interface.  This charge is beyond
what would normally be found in the Schottky compensation for
workfunction mismatch and must be related to a state that exists
at that interface but not found at the tunneling junction
(nanotube - Pt).  In the regime of large tube diameters, it seems
that such states form in the case of NiO and Au - nanotube
contacts, but not in the case of HOPG - nanotube interfaces.
However, to address the transition from large-diameter limit to
small-diameter limit, we have performed first-principles
pseudopotential density-functional calculations of a perfect SWNT
(5,5) supported on a Au (111) surface and on a graphite surface
and compared this to the electronic structure of a single graphene
sheet on top of both substrates (Our model for the large diameter
limit).  For the case of a graphene sheet on gold, we clearly
observe the formation of a simple Schottky barrier dipole Fig.\
\ref{fig3}a.  This result is independent of the relative
orientation of the graphene sheet with respect to the gold
substrate.  The computed graphene work function of 4.42 eV is very
close to the HOPG values and it is smaller than the computed work
function of Au, 5.3 eV.  We have also checked that as we increase
the number of graphene layers, the work function increases with a
small (few meV) dependence on stacking (AB versus AA or random)
sequence.  (A small, negative, contact potential for large
diameter MWNTs on HOPG will result from this difference in work
functions due to the random nature of the tube-layer
stacking).\cite{note}  This is compared to the formation of an
interface potential at the small diameter nanotube - Au contact
shown in Fig.\ \ref{fig3}b.  Because the contact ``distance'' is
not well determined, the graphs compare the corresponding
one-electron effective potential for two distances of the tube
with respect to the substrate (0.22 and 0.34 nm).  In both cases
we see that the small diameter nanotube on Au develops the
opposite barrier height in the vacuum region as that of the
graphene sheet on Au.  Between these two extremes, as the
nanotube's radii are increased, there should be a transition from
one polarity to the other.  This distinct behavior could explain
the observed change of sign in the local interface potential.
Physically, there is a charge distribution at the interface that
leads to either covalent-like or ionic-like bonding states.  The
redistribution of charge density is still observable in the
effective potential even for a tube far from the Au (111) surface
(dashed line in Fig.\ \ref{fig3}b).  This electronic charge
distribution and effective electronic screening is controlled by
the overlap between gold d-states and the outer (interlayer and
surface) tube states. These results assume perfect tubes and that
the outer layer is the only important one as suggested by other
studies on electronic transport in MWNTs.\cite{frank98}
Contributing to this interface charge redistribution is the
opening of small pseudogaps at the Fermi level as a consequence of
breaking the tube-mirror symmetry. Recently, charge density
redistribution has been shown theoretically\cite{ven99} to align
the Fermi-level at the metal-carbon nanotube interface and its
influences on STM spectroscopy have been explored.  We note that
these authors work in the limit of large tube diameter where the
difference in work functions is the key component determining the
contact potential. Further, as argued above, to first order the
overall effect of the nanotube in the tunneling junction cancels
(aside from capacitive effects) unless state-induced Fermi-level
pinning occurs at the interface.

In the case of the HOPG substrate, the bonding is weaker as seen
from the effective potential plotted in Fig.\ \ref{fig3}c for
small diameter tubes.  No Schottky-like dipole is observed for the
particular set of relative orientations chosen.  However, the
vacuum plateau is a little above the zero of energy that we have
taken as the graphitic vacuum level.  This indicates the possible
change of sign in the contact potential with respect to the case
where a single graphene sheet is deposited on graphite.  Moreover,
we can speculate that tube-substrate interactions might play a
significantly different role in formation of interface states when
the tube lies in registry with the substrate. However, this is
still an open question to be resolved in further studies.

In summary, we have related local potentials at the interfaces
between MWNTs and Au, HOPG, and NiO substrates to the formation of
interface states using current transients in rapidly ramped
tunneling spectroscopy.  Sharp variations in these potentials are
observed on Au and HOPG substrates for tube diameters of around 5
nm.  These measurements indicate that the contact electronic
structure is strongly influenced by variations in the electronic
structure of the nanotube as a function of tube diameter.  From
first principles calculations, we demonstrate changes in s-p
hybridization as a function of tube diameter will lead to
differences in the charge distribution at the nano-tube-substrate
interface.  In turn, this results in modifications to the
occupation of interfaces states and leads to the trends observed
in the interface potentials.

The authors would like to acknowledge funding from the following
agencies: DLC (NSF), PMA (SIA/DARPA through the focus center
research program established at RPI). A.R. (support from the
European RTN network contract HPRN-CT-2000-00128 (COMELCAN),
RTD-FET program (SATUNET) and JCyL (VA28/99)).


\begin{references}
\bibitem{short}
``Short'' was determined experimentally by changing filter values
etc. on the microscope and observing the effects on transients
within the system.
\bibitem{ind}
Individual IV spectra were smoothed using 3-point cubic splines so
as to remove ambiguities caused by noise at the very low currents.
Electronic noise in the RHK system we use is approximately 10 fA.
\bibitem{we}
We note that the offset should be the same with and without the
tube and should be equal to the workfunction mismatch of the tip
and the substrate (ie. The contact potential of Au-Pt, etc.)
However, to avoid artifacts due to Gibbs ringing at the digital
voltage steps during ramping, the current transient is measured a
little time after the step edge at which time it has significantly
decreased.  This results in a shift of the curve toward the (0,0)
intercept (steady state) by a constant related to the sample time
delay.  In our system this constant has been determined
independently for any given sample delay.  Using this sampling
time constant the contact potential for the junction is recovered
to within a few percent.
\end{references}
\end{document}